\documentclass[%
pra,
superscriptaddress,
nofootinbib,
amsmath,amssymb,
twocolumn,
floatfix,
longbibliography
]{revtex4-2}

\usepackage[colorlinks=true, linkcolor=purple]{hyperref}
\usepackage{graphicx}%
\usepackage{dcolumn}%
\usepackage{bm}%
\usepackage{stmaryrd} %

\usepackage[utf8]{inputenc}

\allowdisplaybreaks[4]

\graphicspath{{./figures/}}

\makeatletter

\newcommand{\ringplus}{\mathbin{\text{\@ringplus}}}

\newcommand{\@ringplus}{%
  \ooalign{\hidewidth\raise1.3ex\hbox{\tiny$\circ$}\hidewidth\cr$\m@th+$\cr}%
}

\newcommand{\ringminus}{\mathbin{\text{\@ringminus}}}

\newcommand{\@ringminus}{%
  \ooalign{\hidewidth\raise0.9ex\hbox{\tiny$\circ$}\hidewidth\cr$\m@th-$\cr}%
}
\makeatother

\DeclareFontFamily{U}{wncy}{}
\DeclareFontShape{U}{wncy}{m}{n}{<->wncyr10}{}
\DeclareSymbolFont{mcy}{U}{wncy}{m}{n}
\DeclareMathSymbol{\Sh}{\mathord}{mcy}{"58}

\let\originalleft\left
\let\originalright\right
\renewcommand{\left}{\mathopen{}\mathclose\bgroup\originalleft}
\renewcommand{\right}{\aftergroup\egroup\originalright}

\newcommand{\negspace}{\!}

\newcommand{\lsub}[2]{{\protect\vphantom{#1}}_{#2} \negspace {#1}}
\newcommand{\rsub}[2]{{#1} \negspace {\protect\vphantom{#1}}_{#2}}

\newcommand{\ketsub}[2]{\rsub {\ket{#1}} {#2}}
\newcommand{\brasub}[2]{\lsub {\bra{#1}} {#2}}

\newcommand{\pket}[1]{\ketsub{#1} p}
\newcommand{\qket}[1]{\ketsub{#1} q}

\newcommand{\inprod}[2]{\left\langle {#1} | {#2} \right\rangle}

\newcommand{\lsubinprod}[3]{\lsub {\inprod{#1}{#2}} {#3}}

\newcommand{\qleftinprod}[2]{\lsubinprod{#1}{#2} q}

\newcommand{\abs}[1]{\left\lvert{#1}\right\rvert}
\newcommand{\abss}[1]{\lvert{#1}\rvert}

\newcommand{\op}[1]{\hat{#1}}

\newcommand{\opvec}[1]{\op{\vec{#1}}}

\newcommand{\id}[0]{I}

\newcommand{\mat}[1]{\bm{\mathrm{#1}}}

\renewcommand{\vec}[1]{\bm{\mathrm{#1}}}

\newcommand{\ahat}[0]{\hat{a}}
\newcommand{\adagger}[0]{\hat{a}^\dagger}

\newcommand{\floor}[1]{\left\lfloor #1 \right\rfloor}
\newcommand{\ceil}[1]{\left\lceil #1 \right\rceil}

\usepackage{amssymb}
\usepackage{braket}
\usepackage{subfiles}
\usepackage[dvipsnames]{xcolor}
\usepackage{comment}
\usepackage{mathtools}

\date{1 December 2023}

\begin{document}

\title{Biased Gottesman-Kitaev-Preskill repetition code}

\author{Matthew P. Stafford}
\email{matthew.stafford@bristol.ac.uk}
\affiliation{Quantum Engineering Technology Labs, H. H. Wills Physics Laboratory and Department of Electrical and Electronic Engineering, University of Bristol, Bristol, UK.}
\affiliation{Quantum Engineering Centre for Doctoral Training, H. H. Wills Physics Laboratory and Department of Electrical and Electronic Engineering, University of Bristol, Bristol, UK.}

\author{Nicolas C. Menicucci}%
  \email{nicolas.menicucci@rmit.edu.au}
\affiliation{%
 Centre for Quantum Computation and Communication Technology, School of Science, RMIT University, Melbourne, Victoria 3000, Australia}%

\begin{abstract}

Continuous-variable quantum computing architectures based upon the Gottesman-Kitaev-Preskill (GKP) encoding have emerged as a promising candidate because one can achieve fault-tolerance with a probabilistic supply of GKP states and Gaussian operations. Furthermore, by generalising to rectangular-lattice GKP states, a bias can be introduced and exploited through concatenation with qubit codes that show improved performance under biasing. However, these codes (such as the XZZX surface code) still require weight-four stabiliser measurements and have complex decoding requirements to overcome. In this work, we study the code-capacity behaviour of a rectangular-lattice GKP encoding concatenated with a repetition code under an isotropic Gaussian displacement channel. We find a numerical threshold of $\sigma = 0.599$ for the noise's standard deviation, which outperforms the biased GKP planar surface code with a trade-off of increased biasing at the GKP level. This is all achieved with only weight-two stabiliser operators and simple decoding at the qubit level. Furthermore, with moderate levels of bias (aspect ratio~$\leq 2.4$) and nine or fewer data modes, significant reductions in logical error rates can still be achieved for $\sigma \leq 0.3$, opening the possibility of using GKP-biased repetition codes as a simple low-level qubit encoding for further concatenation.
\end{abstract}

\maketitle

\section{Introduction}

Approaches to quantum computation based upon the continuous-variable (CV) formalism have recently seen promising developments within both experiment and theory. On the experimental front, large-scale CV cluster states have been generated optically~\cite{Larsen2019,Asavanant2019,Yokoyama2013,Yoshikawa2016,Chen2014}. Furthermore, single- and two-mode Gaussian operations have also been demonstrated on these platforms as a proof of concept for CV measurement based quantum computation~\cite{Asavanant2021time,larsen2021b}. Theoretical advances have reduced the problem of achieving universal, fault-tolerant computation~\cite{Baragiola2019,Bourassa2021,Tzitrin2021fault} to that of probabilistically generating high quality Gottesman-Kitaev-Preskill (GKP) states~\cite{Gottesman2001}.

In CV quantum computation, an approach to achieving fault-tolerance is through concatenating an inner bosonic code~\cite{Albert2018} with an outer qubit code~\cite{Gottesman2010}. The bosonic code discretises the CV noise to produce effective qubits while taking advantage of redundancy in the Hilbert space to provide error resilience. Examples include cat~\cite{Cochrane1999}, binomial~\cite{Michael2016} and GKP~\cite{Gottesman2001} codes. These encoded modes can then be used as input to a qubit-level code, provided the error rates for the inner layer fall below the qubit code threshold fault-tolerant computation can be achieved~\cite{Menicucci2014}.

Within this approach, GKP codes are particularly attractive because once you have a supply of encoded resource states everything else is Gaussian~\cite{Baragiola2019,Yamasaki2020}, which is considerably easier to perform experimentally. A further benefit for optical architectures is that Gaussian operations can be performed at room temperature, and the overall system can take advantage of the intrinsic scalability offered through integrated photonic platforms~\cite{Bourassa2021,Tzitrin2021fault,Asavanant2021time,larsen2021b}.

The theoretical and experimental progress made so far is yet to coalesce due to the challenge in generating sufficiently high-quality GKP states. Square-lattice GKP states have been experimentally generated on superconducting~\cite{Campagne-Ibarcq2020a} and trapped-ion~\cite{Fluhmann2019,deNeeve2022} platforms. Although recent work has demonstrated GKP error correction beyond break-even~\cite{sivak2023}, state quality currently does not meet the requirements for fault tolerance. In the optical domain, GKP states have not yet been produced, although several generation methods have been proposed~\cite{Hastrup2022,Weigand2018,Su2019,Vasconcelos2010,dahan2023,Eaton2019}. Lessening the experimental burden by reducing both the number of modes and quality of state required to successfully perform GKP error correction is therefore a vital and active area of research. 

Recent work has shown that some codes, such as the original surface code~\cite{KITAEV20032} and, even better, the XZZX surface code~\cite{BonillaAtaides2021}, exhibit improved thresholds under biased noise~\cite{tuckett2019tailoring,Tuckett2020fault}. Therefore, having a bosonic code that produces an effective qubit with biased noise may be advantageous. Rectangular-lattice GKP codes produce biased logical noise when correcting an unbiased channel due to asymmetry in the lattice spacing. Concatenation with the surface code~\cite{Hanggli2020} and the XZZX code~\cite{zhang2023} have both shown an improved threshold under an isotropic Gaussian displacement channel (GDC). However, surface codes have complex decoding requirements and require weight-four stabiliser measurements that may have an undesirable effect on the overall system performance when attempting to scale. 

Just as the Shor code~\cite{Shor1995} is a concatenation of two repetition codes, each protecting against a single type of logical error, we ask, can we take a similar approach with a concatenated GKP code? At the GKP level, we can leverage rectangular-lattice codes to drastically reduce one type of logical error (for example, phase flips) even with a modest lattice aspect ratio (due to super-exponential decay of the tails of the complementary error function). We can then concatenate with a repetition code on the outer layer to correct against the bit-flip errors neglected by the biased GKP code. Not only would this considerably simplify the qubit-level decoder, but it also halves the required connectivity per ancilla for making syndrome measurements to two. Similar approaches have been taken with cat-repetition codes with promising results~\cite{Guillaud2019,Guillaud2021}. If the same applies to a biased GKP-repetition code, we can combine the advantages of Gaussian computation with the simplicity of this code structure. This is the motivation for our study.

In this work, we consider the code-capacity performance of biased GKP-repetition codes under the isotropic GDC parameterised by variance $\sigma^2$. We choose this noise model in order to provide the most direct comparison with the original GKP code, the biased GKP-surface \cite{Hanggli2020} code, and the GKP-XZZX \cite{zhang2023} code, all of which benchmark their performance against this type of noise. (We discuss possible extensions to other noise models in Sec.~\ref{subsec:challenges}.)  The term \emph{code capacity} means we analyse the performance of an ideal GKP code under conditions where the only source of noise is the channel itself. The inner bosonic layer is a rectangular-lattice GKP code that biases the effective qubit to protect against phase-flip errors. Furthermore, the level of bias can be tuned by choosing an appropriate lattice aspect ratio. We then concatenate with a \emph{classical} $n$-qubit repetition code to protect against bit flips. To assess the code performance, we begin by comparing it against a single-mode, square-lattice GKP state. We find the code successfully suppresses the logical error rate, providing the GKP biasing is optimised for both the noise severity (standard deviation~$\sigma$) and number of modes, $n$. We then numerically study the threshold behaviour of the code and find a performance improvement for noise levels below $\sigma \approx 0.599$, remarkably outperforming the GKP-surface code result~\cite{Hanggli2020} when decoding without the analogue methods proposed in~\cite{Fukui2017}. The cost is increased biasing in the GKP lattice. If we restrict the maximum biasing of our code to that considered in~\cite{Hanggli2020}, we find improved code performance for $\sigma < 0.588$. However, with this restriction, error rates cannot be arbitrarily suppressed below the cutoff, and the threshold behaviour is lost. We then shift to analysing the code for resources more applicable to potential small-scale experimental implementations. With up to 9 data modes and modest biasing~$r \leq 2.4$, we can suppress error rates with moderate values of $\sigma < 0.3$ by approximately a factor of~60. With only a few more modes ($n \leq 31$) and a maximum biasing aspect ratio of $r=4$, we can further suppress error rates below those required for fault-tolerance levels when concatenating with a further code~\cite{Noh_2022}. 

The paper is structured as follows. In Section \ref{sec:notation}, we introduce our notation conventions and useful functions. In Section \ref{sec:GKP_code}, we introduce the inner rectangular-lattice GKP code and study its performance under the GDC. Readers familiar with the GKP code structure and GDC are may skip directly to Section \ref{sec:rep_code}, where the outer repetition code is discussed before the two layers are put together to give our GKP-repetition code. Key results and discussion presented in Section \ref{sec:results}. Finally, we summarise our findings and discuss potential extensions for future work in Section \ref{sec:conclusions}. 

\section{Preliminaries}\label{sec:notation}

We begin by reviewing the notation used throughout this paper, following the conventions in~\cite{Gu2009,mensen2021}. Throughout the work, operators acting at the CV, logical-GKP, and repetition-code levels are given by $\hat{O}$, $\Bar{O}$ and~$O_{L}$ respectively. States will likewise be denoted $\ket{\psi}$,  $\ket{\Bar{\psi}}$ and~$\ket{\psi}_{L}$.  Working in units where $\hbar = 1$, the position and momentum quadratures for a single mode of a quantum harmonic oscillator with creation and annihilation operators ($\adagger$, $\ahat$) are given by
\begin{equation}
    \hat{q} = \frac{1}{\sqrt{2}}\left(\ahat + \adagger\right), \quad \hat{p} = \frac{1}{\sqrt{2}i}\left(\ahat - \adagger\right),
\end{equation}
respectively, satisfying the commutation relation $\left[\hat{q},\hat{p}\right] = i \op{\id}$. In these units, the observed vacuum variance for any quadrature is $\sigma^2_{\text{vac}} = 1/2$. Quadrature eigenstates are denoted
\begin{equation}
    \op{q}\qket{s} = s\qket{s}, \quad \op{p}\pket{s} = s\pket{s}.
\end{equation}

We now define the single-mode Gaussian operations that will be used in this paper. Under a Gaussian unitary operation $\hat{G}$, the quadrature operators $\opvec{x} = (\hat{q},\hat{p})^\intercal$ transform according to
\begin{align}\label{eq:Gaussevolution}
\opvec{x} \mapsto \hat{G}^{\dagger} \opvec{x} \hat{G} = \mat{S}_{\hat{G}}\opvec{x} + \vec{c},
\end{align}
where $\mat{S}_{\hat{G}}$ is the symplectic representation of the Heisenberg action of $\hat{G}$ on the quadrature operators~$\opvec x$, along with $\vec{c}$, which represents a linear displacement term. 

\begin{align}
    \hat{X}(g) &\coloneqq e^{-ig\hat{p}},\\
    \hat{Z}(h) &\coloneqq e^{ih\hat{q}},
\end{align}
where their action shifts the quadrature eigenstates by the argument $\hat{X}(g)\qket{s} = \qket{g+s},\ \hat{Z}(h)\pket{t} = \pket{h+t}$. A general displacement operator $\op{D}(g,h)$ is given by%
\footnote{In terms of the usual Glauber displacement operator~\cite{kok2010introduction} ${\op D(\alpha) = e^{\alpha\op{a}^{\dagger} - \alpha^{*}\op{a}}}$, our displacement operator, Eq.~\eqref{eq:displacementop}, is $\op D(g,h) = \op D[(g+ih)/\sqrt{2}]$.}
\begin{equation}\label{eq:displacementop}
\hat{D}(g,h) \coloneqq e^{ih\hat{q} -ig\hat{p}}.
\end{equation}
$\op{D}(g,h)$ translates the quadrature operators in phase space according to
\begin{equation}
    \begin{pmatrix}
    \op{q}\\
    \op{p}
    \end{pmatrix} \mapsto 
    \begin{pmatrix}
    \op{q}\\
    \op{p}
    \end{pmatrix} +
    \begin{pmatrix}
    g\\
    h
    \end{pmatrix}.
\end{equation}

The rotation operator $\op{R}(\theta)$ is a phase delay, which is given by
\begin{align}
    \op{R}(\theta) &\coloneqq e^{i \theta\left(\op{q}^2 + \op{p}^2\right)/2},\\
    \begin{pmatrix}
    \op{q}\\
    \op{p}
    \end{pmatrix} &\mapsto 
    \begin{pmatrix}
    \cos{\theta} & -\sin{\theta}\\
    \sin{\theta} & \cos{\theta}
    \end{pmatrix}
    \begin{pmatrix}
    \op{q}\\
    \op{p}
    \end{pmatrix}
    .
\end{align}
$\op R(\theta)$ rotates the phase-space representation of the state counterclockwise by angle $\theta$ against the mode's direction of normal unitary time evolution (hence, a delay in the phase). For $\theta = \pi/2$ this becomes the Fourier gate~$\hat{F}$. 

The quadrature squeezing operator $\op{S}(s)$, given by
\begin{align}\label{eq:squeeze_operator}
    \op{S}(s) &\coloneqq e^{i\ln{(s)}(\op{q}\op{p} + \op{p}\op{q})/2},\\
    \begin{pmatrix}
    \op{q}\\
    \op{p}
    \end{pmatrix} &\mapsto 
    \begin{pmatrix}
    s^{-1} & 0\\
    0 & s
    \end{pmatrix}
    \begin{pmatrix}
    \op{q}\\
    \op{p}
    \end{pmatrix},
\end{align}
defined in terms of the \emph{squeezing factor}~$s>0$. When $s>1$, this operator
reduces the variance in the position quadrature while simultaneously increasing it in the momentum quadrature, both by a factor of $s$. When $0<s<1$, these effects are reversed. The squeezing factor~$s$ is the exponential of the usual squeezing parameter~\cite{Alexander2016,kok2010introduction}.

Finally, the shear operator is given by
\begin{align}
    \op{P}(t) &\coloneqq e^{it\hat{q}^2 /2},\\
    \begin{pmatrix}
    \op{q}\\
    \op{p}
    \end{pmatrix} &\mapsto 
    \begin{pmatrix}
    1 & 0\\
    t & 1
    \end{pmatrix}
    \begin{pmatrix}
    \op{q}\\
    \op{p}
    \end{pmatrix}.
\end{align}
This operator shears the state's Wigner function along the momentum direction, increasing (resp.,\ decreasing) in momentum for positive (resp.,\ negative) position values. Together, displacements, rotations and either squeezing or shearing are sufficient to enact any single-mode Gaussian operation~\cite{Gu2009}.

A useful phase-space representation of a CV state $\hat{\rho}$ is the Wigner function, defined by
\begin{align}
    W_{\hat{\rho}}(q,p) =  \frac{1}{\pi} \int dy\ \brasub{q + y}{q}\hat{\rho}\ketsub{q - y}{q} e^{2ipy},  
\end{align}
which for a pure state $\op{\rho} = \ket{\psi}\bra{\psi}$ becomes
\begin{equation}\label{eq:wigner}
    W_{\ket{\psi}\bra{\psi}}(q,p) = \frac{1}{\pi}\int dy\ \psi(q + y)\psi^{*}(q - y) e^{2ipy},
\end{equation}
where $\psi(s) = \qleftinprod{s}{\psi}$ is the position-basis wavefunction of the state. For a physical state, $W_{\op{\rho}}(q,p)$ must integrate to one over all phase space. A convenient property of Wigner functions is that, under Gaussian operations, the function transforms according to
\begin{equation}\label{eq:wigner_update}
    W_{\hat{G}\hat{\rho}\hat{G}^\dagger}(\vec{x}) = W_{\hat{\rho}}(\mat{S}_{\hat{G}}^{-1}(\vec{x} - \vec{c})),
\end{equation} 
that is, one just needs to update the coordinates of the function according to the inverse of the Heisenberg action of the operator, Eq.~\eqref{eq:Gaussevolution}.

\section{Rectangular GKP Code}\label{sec:GKP_code}
GKP codes are a way to encode a discrete quantum system within a harmonic oscillator~\cite{Gottesman2001}. These are designed to correct against small displacement errors on the oscillator. Since these errors form an operator basis, the GKP code is capable of protecting to some degree against all errors. Recently, they have shown particular resilience to photon loss~\cite{Noh2019}. Another benefit of the GKP encoding is that it can act as both an error-correcting code and a non-Gaussian resource for universality~\cite{Baragiola2019}. Clifford operations on GKP qubits can all be implemented with Gaussian operations, which are considered easier to perform experimentally. Furthermore, obtaining a magic state~\cite{Bravyi2005} required for universality can be achieved by performing GKP error correction 
on the vacuum~\cite{Baragiola2019}.

In this section, we introduce and discuss the rectangular-lattice GKP codes that form the inner layer of our scheme. We begin by defining ideal square- and rectangular-lattice GKP states before discussing their error-correcting capability under displacements. The isotropic Gaussian displacement channel is then introduced, and we analyse the performance of the single-mode GKP states that produce our effective qubits for concatenation. We find that by changing the shape of the GKP lattice, we bias the logical outcome of the GKP error-correction process. 

\subsection{Square-lattice GKP states}

\begin{figure*}
    \centering
    \includegraphics[width=\linewidth]{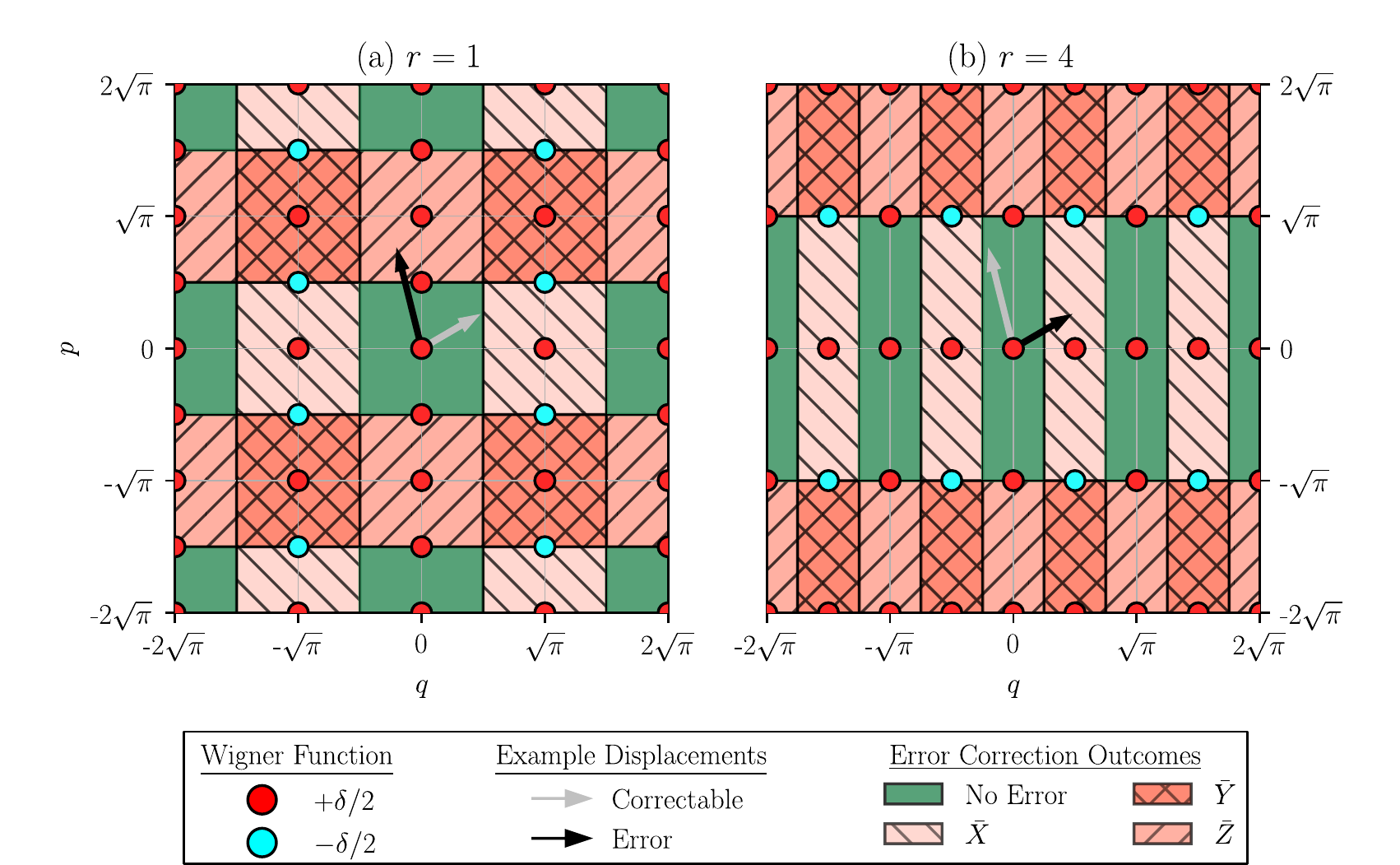}
    \caption{Wigner functions of ideal square ($r = 1$) and rectangular ($r = 4$) GKP $\ket{\Bar{0}}$ states. The functions are periodic arrays of weighted $\delta$-functions. Considering displacement errors, phase space can be partitioned into cells that show the logical outcome of a displacement followed by a round of GKP error correction. Example displacement errors are represented by arrows on the plots. Grey arrows show correctable displacements that terminate in a green unhatched cell, while black displacements result in a logical error and end in one of the hatched cells. Under biasing, the state becomes more resilient to errors in the momentum quadrature at the expense of increased susceptibility to errors in the position quadrature.}
    \label{fig:GKP_Wigner}
\end{figure*}

Ideal, square-lattice GKP-qubit computational-basis states are defined by
\begin{align}\label{eq:GKP0}
    \ket{\Bar{0}} &\coloneqq (2\sqrt{\pi})^{1/2} \sum^{\infty}_{n=-\infty}\qket{2n\sqrt{\pi}},\\
    \label{eq:GKP1}
    \ket{\Bar{1}} &\coloneqq (2\sqrt{\pi})^{1/2} \sum_{n=-\infty}^{\infty}\qket{(2n+1)\sqrt{\pi}},
\end{align}
they are an infinite superposition of position quadrature eigenstates spaced $2\sqrt{\pi}$ apart. These states cannot be normalised and are therefore unphysical. The factor of $(2\sqrt{\pi})^{1/2}$ is included here to simplify the transformations to the conjugate basis. Due to their periodicity, they---and thus all GKP code states---are stabilised by the group generated by $\mathcal{S} = \langle \hat{X}(2\sqrt{\pi}),\: \hat{Z}(2\sqrt{\pi})\rangle$.  Logical Pauli operators are given by the displacements $\Bar{X} = \op{X}\left(\sqrt{\pi}\right),\: \Bar{Z}= \op{Z}\left(\sqrt{\pi}\right)$---i.e.,~by halving the magnitude of the stabiliser operators. An arbitrary pure GKP qubit is given by $\ket{\Bar{\psi}} = \alpha\ket{\Bar{0}} + \beta\ket{\Bar{1}}$, with $\abs{\alpha}^2 + \abs{\beta}^2 = 1$. 

The position-basis wavefunctions for the logical states $\psi_j(s) = \lsub{\braket{s | \Bar{j}}}{q}$ for $\ j \in \{0,1\}$ are given by
\begin{equation}\label{eq:GKP_square_pos_wavefunction}
    \psi_{j}(s) = (2\sqrt{\pi})^{1/2} \sum^{\infty}_{n = -\infty} \delta(s + \sqrt{\pi}(2n + j)),
\end{equation}
these are Dirac combs of period $2\sqrt{\pi}$, with $j$ encoding the half-period shift required for $\Bar{\ket{1}}$. By taking the Fourier transform of Eq.~\eqref{eq:GKP_square_pos_wavefunction} we obtain the momentum basis wavefunctions~$\tilde\psi_j(s) = \lsub{\braket{s | \Bar{j}}}{p}$, which evaluate to%
\footnote{We distinguish position- from momentum-space wavefunctions solely by the tilde---i.e.,~$\psi(\cdot)$ versus $\tilde\psi(\cdot)$, respectively---and irrespective of the letter used for the function's argument, which is kept generic.}%
\begin{equation}\label{eq:GKP_sqare_mtm_wavefuntion}
    \tilde{\psi}_{j}(s) = (\sqrt{\pi})^{1/2} \sum^{\infty}_{n = -\infty} (-1)^{jn}\delta(s + n\sqrt{\pi}),
\end{equation}
where the comb period has halved to $\sqrt{\pi}$ and now has an alternating phase term depending on the logical content of the state. The prefactor of this function also differs from that in Eq.~\eqref{eq:GKP_square_pos_wavefunction} by a factor of $\sqrt 2$, which preserves the norm of the state.

Substituting the $\ket{\Bar{0}}$ wavefunction from Eq.~\eqref{eq:GKP_square_pos_wavefunction} into Eq.~\eqref{eq:wigner}, we obtain the square-lattice Wigner function
\begin{equation}
W_{\ket{\Bar{0}}\bra{\Bar{0}}}(q,p) = \frac{1}{2} \sum_{m,n}(-1)^{mn}\delta\left(q - n\sqrt{\pi}\right)\delta\left(p - m\frac{\sqrt{\pi}}{2}\right),
\end{equation}
which is an infinite, periodic array of weighted $\delta$-functions, displayed in Figure \ref{fig:GKP_Wigner}(a).

\subsection{Generalising to rectangular lattices}
\label{subsec:reclattice}

While square-lattice GKP states are most widely studied due to their balanced protection in each quadrature, we are free to define alternative-lattice grid states as long as the stabiliser generators commute~\cite{Gottesman2001}. Keeping the direction of each displacement but allowing the magnitude to be different, we have 
\begin{equation}
    \left[\hat{X}(g),\hat{Z}(h)\right] = 0 \iff gh = 2\pi d,
\end{equation}
where $d$ specifies the dimension of the logical subspace contained within the code.\footnote{In this work, we focus on $d=2$, but we show the general relation for completeness.} For qubits ($d=2$), the generalisation from the square lattice to a rectangular one results from letting $g = 2\sqrt{\pi /r}$ and $h = 2\sqrt{\pi r}$ for $r \in \mathbb{R} > 0$. The stabiliser generators become $\langle \hat{X}(2\sqrt{\pi/r}),\hat{Z}(2\sqrt{\pi r})\rangle$ and the logical operators are given by $\op{X}(\sqrt{\pi/r}),\ \op{Z}(\sqrt{\pi r})$. Setting $r=1$ recovers the square-lattice GKP states. The position- and momentum-space wavefunctions, Eqs.~\eqref{eq:GKP_square_pos_wavefunction} and~\eqref{eq:GKP_sqare_mtm_wavefuntion}, generalise to
\begin{align}
    \psi_{j,r}(s) &=  (2\sqrt{\pi / r})^{1/2} \sum^{\infty}_{n = -\infty} \delta(s + \sqrt{\pi /r}(2n + j)),\\
    \tilde{\psi}_{j,r}(s) &=  (\sqrt{\pi r})^{1/2} \sum^{\infty}_{n = -\infty} (-1)^{jn}\delta(s + n\sqrt{\pi r}).
\end{align}
The Wigner function for the rectangular $\ket{\Bar{0}}$ state parameterised by $r$ is given by
\begin{equation}\label{eq:biased_wigner}
W(q,p) = \sum_{m,n}\frac{(-1)^ {mn}}{2}\delta\left(q - n\sqrt{\frac{\pi}{r}}\right)\delta\left(p - m\frac{\sqrt{\pi r}}{2}\right),
\end{equation}
shown in Figure~\ref{fig:GKP_Wigner}(b) for $r=4$. The periodicity of the state has gone from $2\sqrt{\pi}$ in both quadratures to $2\sqrt{\pi / r}$ in the position quadrature and $2\sqrt{\pi r}$ in the momentum quadrature. In this sense, $r$ can be interpreted as the aspect ratio of the lattice.

\begin{figure}
    \centering
    \includegraphics[width=\linewidth]{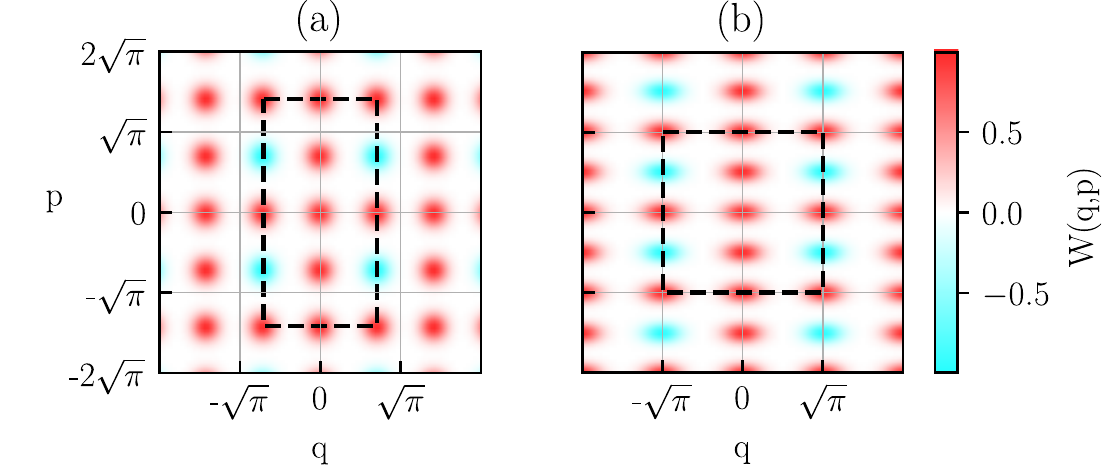}
    \caption{(a) Wigner function for an $r=2$ rectangular $\ket{\Bar{0}}$ GKP state after undergoing Guassian displacement noise, Eq.~\eqref{eq:isogauss}, with $\sigma=0.2$, corresponding to GKP squeezing of approximately $11\text{dB}$. Each of the $\delta$ peaks in the ideal Wigner function, Eq.~\eqref{eq:biased_wigner}, have now been blurred into a narrow Gaussian through convolution with the noise channel. A unit cell of the lattice is enclosed within the dashed rectangle, over which the Wigner function integrates to one. (b) A square-lattice state that has undergone a biased noise channel with $\sigma_q = \sigma \sqrt{r}$, $\sigma_p = \sigma / \sqrt{r}$. While these two states are physically distinct, their GKP correction outcomes are identical.}
    \label{fig:noisy_wig_function}
\end{figure}

An alternative way to consider rectangular-lattice states is by squeezing the corresponding ideal square-lattice state. As the squeezing operator, Eq.~\eqref{eq:squeeze_operator}, is Gaussian, we can use the Wigner function update rule, Eq.~\eqref{eq:wigner_update}. Applying $\hat{S}(\sqrt{r})$ to a square-lattice $\ket{\Bar{0}}$ state results in the rectangular-lattice Wigner function, Eq.~\eqref{eq:biased_wigner}. Here, we stress that this interpretation only applies to ideal GKP states. If we attempted this method with an approximate square-lattice GKP state, we would not obtain the equivalent approximate rectangular lattice state because the Gaussian peaks (which replace the ideal $\delta$-functions in the Wigner function) would also be distorted by the squeezing operation. An example is shown in Figure~\ref{fig:noisy_wig_function}.

\subsection{Correcting displacement errors}\label{sec:correction_process}

While ideal GKP states are unphysical, it is still useful to study their error-correction capability in a \emph{code-capacity model}. Specifically, this is the error model in which all states, operations and measurements are assumed to be ideal. This shows us in principle whether a code can suppress errors under a certain noise channel.%
\footnote{The understanding remains that physical states and other imperfections will reduce this suppression in practice, but the value of the code-capacity model is that it is the in-principle best that can be done with the code, and this is useful as an initial analysis of the code's potential.} %
The effect of the GKP decoding process is to discretise the continuous noise in each quadrature to a binary outcome. Correction always returns us to the original GKP codespace with either the identity (in the case of successful correction) or a logical Pauli operator applied to the state~\cite{Gottesman2001}. To perform this correction in practice, one can either implement Steane-style~\cite{Glancy2006,Steane1997} or Knill-style (i.e.~teleportation-based)~\cite{Walshe2020,Knill2005} error correction.

The GKP encoding has an inbuilt error-correction capability against small displacements, schematically shown in Figure~\ref{fig:GKP_Wigner}. We begin by noting that any ideal GKP qubit only has Hilbert-space support on $\{\qket{n\sqrt{\pi/r}} \mid n\in \mathbb{Z}\}$ in position and $\{\pket{m\sqrt{\pi r}} \mid m\in \mathbb{Z}\}$ in momentum.
Therefore, taking quadrature measurements modulo $\sqrt{\pi /r}$ for position and $\sqrt{\pi r}$ for momentum will always return zero. Let $\ket{\Bar{\psi}}$ undergo an arbitrary displacement $\op{D}(g,h)$. Disregarding global phase, the corrupted state is $\op{Z}(h)\op{X}(g)\ket{\Bar{\psi}}$. Measuring the quadratures will return $g = n\sqrt{\pi / r} + \delta_{q}$, $h = m\sqrt{\pi r} + \delta_{p}$, where we have split the result into the codespace and remainder terms, the latter satisfying $\abs{\delta_{q}} \leq \sqrt{\pi / 4 r}$, $\abs{\delta_{p}} \leq \sqrt{\pi r / 4}$. In practice, we only have access to the remainder and therefore only have information about how far away the state is from the codespace.%

To attempt to correct the displacement, we implement the typical nearest-lattice-point decoder~\cite{Gottesman2001}, which assumes that smaller displacements are more likely to occur. We apply the correction $\op{X}(-\delta_{q})\op{Z}(-\delta_{p})$ that brings the state back to the centre of the nearest cell in Figure~\ref{fig:GKP_Wigner}. If $n$ and $m$ are both zero, then the correction returns the system to the original state and corrects the displacement error, indicated by the solid green region around the origin. If $m$ or $n$ is odd, then the correction operation does not return the system to the original state. Instead $\ket{\Bar{\psi}}$ is displaced to the centre of a hatched red cell, and a logical error is introduced. Using this decoder, we see that the rectangular-lattice codes can tolerate displacement errors of magnitude
\begin{equation}
    \lvert g \rvert < \sqrt{\pi /4r}, \quad \lvert h \rvert < \sqrt{\pi r/4}.
\end{equation}
Example displacements are shown in Figure~\ref{fig:GKP_Wigner}, represented as arrows from the origin. If the displacement lies within a solid green cell, the GKP correction process will succeed, represented here by grey arrows. For the cases where the displacement is too large, shown here by black arrows, the state is moved out of the central region into one of the hatched red cells. The correction displaces the state to the centre of the cell in which the state now lies, resulting in a logical error being applied. For the $r=1$ state in Figure~\ref{fig:GKP_Wigner}(a), a $\Bar{Z}$ error occurs. For the $r=4$ state in Figure~\ref{fig:GKP_Wigner}(b), failure results in the application of $\Bar{X}$. Here, we see that choosing an alternate lattice results in different error-correction outcomes. The $r=4$ state is now much more resistant to $\hat{Z}$ shift errors at the expense of becoming less tolerant to $\hat{X}$ displacements.

Finally, we consider the case where $n$ and~$m$ are both even. Even though the original displacement is much larger than the correction operation, snapping back to the lattice results in even powers of $\Bar{X}$ and $\Bar{Z}$ being applied. These are stabilisers. Therefore, the code can correct for some large displacements, shown by the green regions of phase space not at the origin. For the square-lattice GKP code under typical noise models, these large displacements have a negligible contribution to the outcome probabilities and are therefore discarded. However, when considering rectangular-lattice states, we need to take these effects into account.

\subsection{Noise model: isotropic Gaussian displacement channel}\label{sec:noise_model}

In this work, we consider the isotropic Gaussian displacement channel~(GDC) as our noise model, given by
\begin{align}\label{eq:isogauss}
    \mathcal{N}_f &= \int_{\mathbb{R}^2} dg\, dh\, f(g,h)\hat{D}(g,h) \odot \hat{D}^{\dagger}(g,h),\\
    f(g,h) &= \frac{1}{2\pi\sigma^2}e^{-(g^2 + h^2)/2\sigma^2},\label{eq:noise_double}
\end{align}
where $\hat{D}(g,h)$ is the displacement operator, Eq.~\eqref{eq:displacementop}, whose the displacement amounts~$(g,h)$ are randomly drawn from the Gaussian distribution~$f(g,h)$. The channel output, $\mathcal N_f(\op\rho)$, is a weighted average of these displacements applied to the input state~$\op \rho$. Under this channel, the input state suffers from an unknown but definite displacement. The severity of the noise channel is characterised by the variance~$\sigma^2$. The effect of this channel on~$\ket{\Bar{0}}$ is to convolve its ideal Wigner function [Eq.~\eqref{eq:biased_wigner} with $j=0$] with the noise distribution, Eq.~\eqref{eq:noise_double}, resulting in each $\delta$ peak blurring into an isotropic Gaussian with variance $\sigma^2$. An example of an $r=2$ state under a GDC of $\sigma=0.2$ is shown in Figure~\ref{fig:noisy_wig_function}(a). In accordance with the literature, we can characterise the noise on each of the spikes relative to the vacuum using the GKP squeezing factor~\cite{Hanggli2020}
\begin{align}\label{eq:GKP_squeezing}
    s_{\text{GKP}} &= -10\log_{10}\left(\frac{\sigma^2}{\sigma_{\text{vac}}^2}\right)
    ,
\intertext{which can be inverted to give}
    \sigma &= 10^{-s_{\text{GKP}}/20} \sigma_{\text{vac}}
    .
\end{align}
Recall that in this work, $\sigma_{\text{vac}}^2 = 1/2$. For reference, squeezing factors~$s_{\text{GKP}}$ of 0~dB, 3~dB, 6~dB, and 9~dB correspond to standard deviations~$\sigma$ of 0.707%
, 0.501, 0.354, and 0.251, respectively. The negative sign in the definition sets the convention that $\sigma$ decreases for increasing~$s_{\text{GKP}}$.

While the Wigner function is no longer singular, the state is still unphysical because it retains its original periodicity over all phase space. One cannot arrive at an approximate physical GKP state from blurring an ideal state, but such states are still a useful model of noisy GKP states in certain applications~\cite{Menicucci2014,Noh2020,mensen2021,Bourassa2021}. The noisy Wigner function is normalised such that it integrates to one over a unit cell of the lattice, enclosed by the dashed region in Figure~\ref{fig:noisy_wig_function}.

\begin{figure*}
    \centering
    \includegraphics[width=\linewidth]{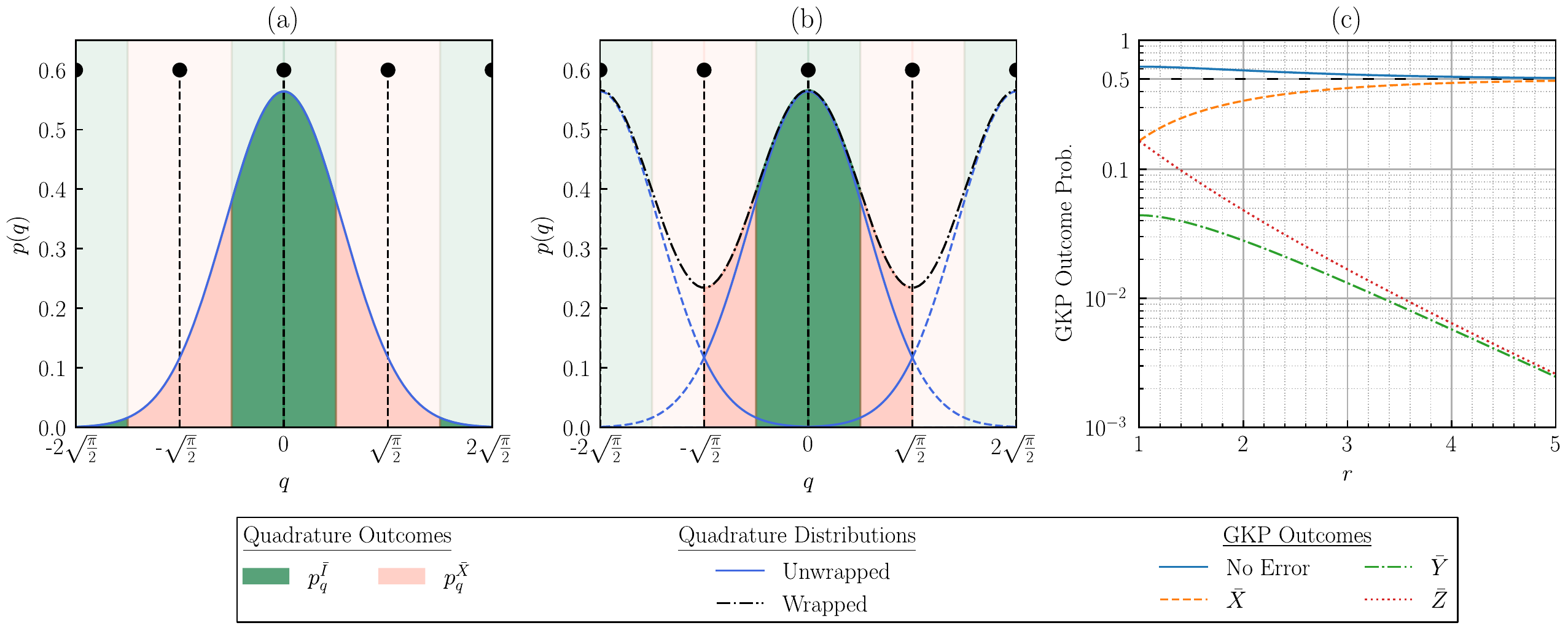}
    \caption{GKP error-correction outcomes for rectangular-lattice states under the Gaussian displacement channel, Eq.~\eqref{eq:isogauss}. Subfigures (a) and~(b) show the position quadrature of an $r=2$ GKP state under the GDC with $\sigma = 1/\sqrt{2}$, corresponding to $0~\mathrm{dB}$ of GKP squeezing. The noise distribution in (a) is binned according to the GKP correction outcome, with green (darker) indicating successful correction and red (lighter), the introduction of an $\Bar{X}$ error. The ideal codespace support at every $\sqrt{\pi /2 }$ is represented by the dashed black stems. To obtain the overall outcome probabilities in (a), the contributions from all bins must be summed, which extend into the tails of the distribution. This is achieved by wrapping the distribution---i.e.,~summing all periodically displaced replicas of the original distribution---with a period $T_q = \sqrt{2 \pi}$, as shown in~(b). Wrapping the distribution yields Eq.~\eqref{eq:wrapped}, shown by the dot-dashed black curve. All the outcome information is now contained within the domain $\big[-\sqrt{\pi/2},\sqrt{\pi/2}\big)$, shown schematically by the darker regions in~(b). Integrating Eq.~\eqref{eq:wrapped} over these new bins produces the quadrature outcome probabilities given by Eqs.~\eqref{eq:id} and~\eqref{eq:err}. Subfigure (c) shows how the overall GKP qubit-level outcomes, given by Eqs.~\eqref{eq:gkpid}--\eqref{eq:gkpz}, change with the biasing: increasing $r$ has the effect of suppressing $\Bar{Z}$ and $\Bar{Y}$ errors at the expense of increasing the $\Bar{X}$ error rate, therefore biasing the qubit. Note that without further concatenation, biasing the GKP code reduces its error correction performance, indicated by the decrease in the identity (no-error) outcome probability towards 0.5.}
    \label{fig:GKP_quadratures}
\end{figure*}

To analyse the performance of the GKP code under this channel we can abstract away the correction process and only consider $f(g,h)$. As this distribution is separable we are free to study the quadratures independently. Each quadrature has a displacement probability given by
\begin{equation}\label{eq:quadraturedist}
    f(u) = \frac{1}{\sqrt{2\pi\sigma^2}}e^{-u^2/2\sigma^2},
\end{equation}
and $f(g,h) = f(g) f(h)$. Recalling the error-correction process in Section \ref{sec:correction_process}, we note that the outcome of the correction process is to snap the state to the nearest lattice point, located every $\sqrt{\pi /r}$ for the position quadrature in a rectangular code. This results in either a successful correction or $\Bar{X}$ being applied. We bin the distribution, Eq.~\eqref{eq:quadraturedist}, about each lattice point according to the correction outcome. Figure~\ref{fig:GKP_quadratures}(a) shows the binned outcomes for the position quadrature of an $r=2$ code with $\sigma= 1/\sqrt{2}$. These bins are centered around $n \sqrt{\pi / r},\: n \in \mathbb{Z}$, which is the support of the ideal codespace.

When implementing GKP error correction, many different outcomes of measuring $\op q$ or $\op p$ all result in the same remainder~$\delta_q$ or~$\delta_p$, respectively. These remainders are the quantities used in position-space GKP error correction~\cite{Gottesman2001}. Thus, the probability density of any given value of a remainder is actually a sum of all physical outcomes that have that value as their remainder. 
To obtain this probability density, we must 
wrap the distribution of the physical outcomes onto itself, with a period $T = 2\sqrt{\pi /r}$ centered about the origin, and sum the contributions. The resulting distribution is shown in Figure~\ref{fig:GKP_quadratures}(b) for the position quadrature. In this case, the distribution is wrapped from $-T/2$ to $T/2$. The wrapping process is equivalent to summing a Gaussian pulse train of period $T$, shown by dashed lines in Figure~\ref{fig:GKP_quadratures}(b). The wrapped probability distribution is 
\begin{equation}
    f_{\text{w}}(u,T) = \sum_{k \in \mathbb{Z}} f(u + kT), \quad u \in \left[-T/2,\ T/2\right),
\end{equation}
where we now have all the outcome information within the domain $\left[-T/2,\ T/2\right)$. When $f$ is a zero-mean Gaussian distribution of the form of Eq.~\eqref{eq:quadraturedist}, $f_{\text{w}}$ becomes~\cite{mensen2021}
\begin{equation}\label{eq:wrapped}
    f_{\text{w}}(u,\sigma,T) = \frac{1}{T}\vartheta\left(\frac{u}{T},\frac{2\pi i \sigma^2}{T^2}\right)
\end{equation}
where $\vartheta(z,\tau)$ is the Jacobi theta function of the third kind given by
\begin{equation}
    \vartheta(z,\tau) = \sum_{m \in \mathbb{Z}} \exp{\left[2\pi i \left(\frac{m^2\tau}{2} + mz\right)\right]}.
\end{equation}
The wrapped distribution, Eq.~\eqref{eq:wrapped}, is shown in black on Figure~\ref{fig:GKP_quadratures}(b) for $r=2, \: T= \sqrt{2\pi}$. This is the outcome probability distribution for the noisy GKP mode. It is normalised such that
\begin{equation}
    \int^{T/2}_{-T/2} du \ f_{\text{w}}(u,\sigma ,T) = 1.
\end{equation}
To obtain the individual quadrature outcome probabilities, we integrate Eq.~\eqref{eq:wrapped} over the bins 
\begin{align}
    p^{\Bar{I}}_{q|p} &\coloneqq 2 \int_{0}^{T/4} du \ f_{\text{w}}(\delta_{q|p},\sigma,T),\label{eq:id}\\
    p^{\Bar{E}}_{q|p} = 1 - p^{\Bar{I}}_{q|p} &= 2 \int_{T/4}^{T/2} du \ f_{\text{w}}(\delta_{q|p},\sigma,T),\label{eq:err}
    \end{align}
where the notation $q|p$ indicates two alternatives (in this case, $q$ or $p$), $p^{\Bar{I}}_{q|p}$ is the probability of successful correction, and $p^{\Bar{E}}_{q|p}$ is the probability that a logical error~$\Bar E$ has occurred ($\Bar{X}|\Bar{Z}$, respectively, for $q|p$). Since the distribution is symmetric about the origin, we need only integrate over half of each relevant bin and then double the outcome.

For rectangular GKP codes, the wrapping periods for each quadrature are given by the stabiliser operators. When biased by $r$, we have $T_{q} = 2\sqrt{\pi/r}$ and $T_{p} = 2\sqrt{\pi r}$. The period changes according to $r$, which, in turn, impacts the error probability for each quadrature. For $r=2$, $\sigma = 1 / \sqrt{2}$, as shown in Figure~\ref{fig:GKP_quadratures}, the position-error rate is $p^{\Bar{X}}_q = 0.37$, and the momentum-error rate is $p^{\Bar{Z}}_p = 0.08$, clearly showing a bias towards bit-flip errors. Contrast this with the square-lattice state under the same channel, where the error rate is $0.21$ in each quadrature.

To approximate the rate of error suppression in the momentum quadrature when increasing the aspect ratio of the lattice we first note that the Jacobi theta function in Eq.~\eqref{eq:wrapped} can be well approximated by a Gaussian over a single period $\left[-T/2,  \ T/2\right)$ as $T$ is increased for a fixed~$\sigma$. This is due to there being less of a `tail' in the original Gaussian of Eq.~\eqref{eq:quadraturedist} to wrap. Increasing the biasing~$r$ does exactly this: the momentum wrapping period is scaled with a factor of $\sqrt{r}$ while the position period gets scaled by $1/\sqrt{r}$. Approximating $f_\text{w}$ as a Gaussian in Eq.~\eqref{eq:err} and then taking the upper limit to be $\infty$ (which over-approximates the error) results in the error probability being given by the complementary error function (erfc) evaluated at $T_{p}/4 = \sqrt{\pi r}/2$. Increasing the biasing $r$ for a fixed $\sigma$ simply increases the argument of erfc, which decays super-exponentially.

The individual quadrature results from Eqs.~\eqref{eq:id} and~\eqref{eq:err} are then combined to give GKP qubit-level outcome probabilities. As the quadrature results are independent, the logical outcome rates are simply the intersection of individual quadrature outcomes
\begin{align}
    P_{\text{GKP}}\left[\Bar{I}\right] &= p_{q}^{\Bar{I}}p_{p}^{\Bar{I}},\label{eq:gkpid} \\
    P_{\text{GKP}}\left[\Bar{X}\right] &= p_{q}^{\Bar{X}}p_{p}^{\Bar{I}},\\
    P_{\text{GKP}}\left[\Bar{Y}\right] &= p_{q}^{\Bar{X}}p_{p}^{\Bar{Z}},\\
    P_{\text{GKP}}\left[\Bar{Z}\right] &= p_{q}^{\Bar{I}}p_{p}^{\Bar{Z}}.\label{eq:gkpz}
\end{align}
Under this model, the overall noise and correction process on an encoded GKP qubit can be treated as an independent noise channel
\begin{align}\label{eq:channel}
    \Bar{\rho} &\mapsto \phantom{+}P_{\text{GKP}}\left[\Bar{I}\right]\Bar{\rho} + P_{\text{GKP}}\left[\Bar{X}\right]\Bar{X}\Bar{\rho}\Bar{X}\nonumber\\ &\phantom{\mapsto}+P_{\text{GKP}}\left[\Bar{Y}\right]\Bar{Y}\Bar{\rho}\Bar{Y} + P_{\text{GKP}}\left[\Bar{Z}\right]\Bar{Z}\Bar{\rho}\Bar{Z}.
\end{align}
Figure~\ref{fig:GKP_quadratures}(c) shows the outcome probabilities with increasing~$r$. As the lattice aspect ratio $r$ is increased, $\Bar{Z}$ and $\Bar{Y}$ errors are suppressed at the expense of increasing the $\Bar{X}$ error rate, resulting in a qubit suffering from biased noise. 

The level of bias can be tuned by choosing an appropriate aspect ratio, which we will find beneficial when concatenating with the repetition code. In the limit of infinite biasing, $r \rightarrow \infty$, we see that the $\Bar{Z}$ and $\Bar{Y}$ errors are completely suppressed, while the identity~$\Bar{I}$ and bit-flip~$\Bar{X}$ outcomes approach $0.5$ from above and below, respectively. The effective channel, Eq.~\eqref{eq:channel}, therefore approaches that of a bit-flip channel with equal probability. We note that when only considering a single mode, moving to a rectangular-lattice reduces the overall success probability as seen by the solid blue line---it approaches $0.5$ from above with increasing $r$. 

\subsection{Formal equivalence to a biased noise channel with a square-lattice GKP code}

So far, we have described our system as a rectangular-lattice GKP code subject to an unbiased noise channel. An alternative way to bias a GKP qubit is through a biased GDC acting on a square-lattice code. While these are two distinct physical systems, we now show their GKP error correction outcomes are identical. To see this, consider a rectangular code whose wrapping periods are given by $T_{q} = 2\sqrt{\pi / r}$ and $T_{p} = \sqrt{\pi r}$ for the position and momentum quadratures, respectively. The equivalence can be shown by a simple change of variables. As an example, the successful outcome probability in the position quadrature is given by
\begin{align}
    p_{q}^{\Bar{I}} &= 2\int_{0}^{T_{q}/4} du \ \frac{1}{T_{q}} \vartheta\left(\frac{u}{T_{q}},
    \frac{2\pi i \sigma^2}{T_{q}^{2}}\right),\\
    &= 2\int_{0}^{T/4\sqrt{r}} du \ \frac{\sqrt{r}}{T} \vartheta\left(\frac{u\sqrt{r}}{T},\frac{2\pi i \sigma^2 r}{T^{2}}\right),
\end{align}
where we have substituted in the square-lattice period $T = 2\sqrt{\pi}$ in the second line. We now change integration variables to ${u^\prime = u\sqrt{r}}$ to obtain
\begin{equation}
    p_{q}^{\Bar{I}} = 2\int_{0}^{T/4} du^{\prime} \ \frac{1}{T} \vartheta\left(\frac{u^{\prime}}{T},\frac{2\pi i \left(\sigma \sqrt{r}\right)^2}{T^{2}}\right).
\end{equation}
The integrand is simply~$f_{\text{w}}\left(u^{\prime},\sigma\sqrt{r},T\right)$, which is exactly what is obtained when the square-lattice position quadrature is subject to Gaussian displacement noise with variance $\sigma_{q}^{2} = \sigma^2 r$. Equivalent results for the $\Bar{X}$ outcome are obtained using the same method. For the momentum quadrature, the noise variance becomes $\sigma_{p}^{2} = \sigma^2 / r$. Therefore, in the code-capacity model, subjecting a rectangular-lattice GKP code to unbiased noise has the exact same outcome as a square-lattice code under a biased GDC.

Wigner functions for both cases are shown in Figure~\ref{fig:noisy_wig_function} for the $\sigma=0.2$,\: $r=2$ case. In (a), the rectangular-lattice state undergoes an isotropic GDC. The equivalent interpretation with a biased channel is shown in (b)---this state has a square lattice, but the Gaussian peaks themselves are now biased, showing increased position quadrature noise. While these are very different physical systems in terms of the state and noise channel, under GKP error correction, they are functionally equivalent. Furthermore, one can map between the two states by applying the squeezing operator, Eq.~\eqref{eq:squeeze_operator}, and its inverse. We can also now see why one cannot obtain physical rectangular lattice GKP states from squeezing a noisy square states (Cf.~the end of Sec.~\ref{subsec:reclattice}): The noise distribution would also scale with the squeezing, yielding no change in the GKP correction process outcomes. 

This completes the discussion of the rectangular GKP code, which is the inner layer of our concatenated code. We have seen that under the Gaussian displacement channel, we can both discretise and bias the CV level noise by employing an appropriately shaped GKP lattice. Each mode can now be viewed as an effective qubit suffering from biased noise. In the next section, we turn our attention to the qubit repetition code for concatenation to further suppress logical error rates. 

\section{Concatenated GKP-Repetition Code}\label{sec:rep_code}

Now that we have effective qubits from the GKP code, we can attempt to further suppress error rates through concatenation with an appropriate qubit code. Normally, a code that can provide protection against both bit- and phase-flip errors is chosen to suppress the overall error rate (as in Ref.~\cite{Hanggli2020}). We have seen, however, that by choosing an appropriate GKP lattice, we can bias the noise such that the mode is much more resilient to one type of error---in our case, phase flips. Just as the Shor code~\cite{Shor1995} is a concatenation of two classical repetition codes, each protecting against a single type of error, here we concatenate the biased GKP code, which already protects against phase flips through the lattice asymmetry, with a classical repetition code to handle the residual bit-flip errors.
In this section we review the repetition code and study its performance. 

The $n$-qubit bit-flip repetition code is given by the encoding
\begin{equation}
    \ket{0}_L = \ket{\Bar{0}}^{\otimes n}, \quad \ket{1}_L = \ket{\Bar{1}}^{\otimes n}.
\end{equation}
Eventually we will use GKP states to represent $\ket{\Bar{0}}$ and~$\ket{\Bar{1}}$, but the calculations here apply to any type of qubit subject to fixed Pauli noise.
The stabiliser generators for the code are given by the set of pairwise $\Bar{Z}$ operators acting on adjacent modes, $\mathcal{S} = \langle \Bar{Z}_1\Bar{Z}_2, \Bar{Z}_2\Bar{Z}_3,...,\Bar{Z}_{n-1}\Bar{Z}_n\rangle$.

The decoding process is a simple majority vote, so this code can detect up to~$\ceil{(n-1)/2}$ bit-flip errors and correct up to $\floor{(n-1)/2}$ bit-flip errors. Generally, one prefers odd values of~$n$ because then all detected errors can be corrected. (For even~$n$, exactly $n/2$ flips are detectable as an error but cannot be corrected since either code state is equally likely.) 
For instance, both an $n=3$ and $n=4$ code can only correct at most a single bit flip, although the latter can detect (but not correct) two flips. Therefore, we will only consider odd-$n$ codes. The code also does not have built-in protection against any phase-flip errors since they always commute with all elements of $\mathcal{S}$, and this needs to be accounted for. We now proceed to analyse each error independently before combining to give the overall code performance.

Each data qubit may independently suffer a random bit-flip error, with error rate given by $p_{\Bar{X}}$, with $p_{\Bar{I}} = 1 - p_{\Bar{X}}$ (identical rate for all data qubits).  Given the error probabilities for a Pauli channel are independent and identically distributed (i.i.d.) Bernoulli trials,\ the exact number of bit flips, $j$, on $n$~total qubits, is a random variable following a binomial distribution, $j \sim \mathrm B(n,p_{\Bar{X}})$. The probability that the number of bit flips, denoted $\abss{\Bar{X}}$, is exactly $j$ is thus
\begin{align}
    \Pr\left(\abss{\Bar{X}} = j\right) 
    = \binom {n} {j} p_{\Bar{X}}^j\left(1-p_{\Bar{X}}\right)^{n-j}.
\end{align}
The repetition code can identify and correct for any combination of bit-flip errors on up to $k \coloneqq \floor{(n-1)/2}$ qubits. Therefore, 
the code's bit-flip success probability~$P(\abss{\Bar{X}} \leq k)$ 
is given by the binomial distribution's cumulative distribution function~(CDF) evaluated at $k$,
\begin{equation}
\label{eq:PXlesskbinomial}
   \Pr(\abss{\Bar{X}} \leq k) = F(k;n, p_{\Bar{X}}) = \sum^{k}_{j=0} \binom {n} {j} p_{\Bar{X}}^j\left(1-p_{\Bar{X}}\right)^{n-j},
\end{equation}
which is just a partial binomial sum containing all combinations of outcomes where at most $\floor{(n-1)/2}$ qubits have suffered a bit flip. This CDF, Eq.~\eqref{eq:PXlesskbinomial}, is well known to evaluate to
\begin{align}\label{eq:bitflip_rep}
    \Pr(\abss{\Bar{X}} \leq k) &= I_{1-p_{\Bar{X}}} (n-k,k+1),
\end{align}
where 
\begin{align}\label{eq:incompletebeta}
    I_{x}(a,b) &\coloneqq \frac{B(x;a,b)}{B(1; a,b)}
\end{align}
is the 
regularised incomplete beta function, with 
\begin{align}
    B(x;a,b) &\coloneqq \int_{0}^{x} t^{a-1}(1-t)^{b-1} \ dt
\end{align}
being
the incomplete beta function, which becomes the complete beta function for $x=1$. Thus, Eq.~\eqref{eq:bitflip_rep} is the success probability for correcting a bit-flip error,
where we recall $k = \floor{(n-1)/2}$%
. The bit-flip error rate is then just given by $1 - \Pr\left(\abss{\Bar{X}} \leq k\right)$. The code admits a threshold at $p_{\Bar{X}} = 0.5$ for noise models that only consider bit-flip errors. We recall from Section~\ref{sec:noise_model} that any biased GKP state under the GDC has a bit-flip error rate~$p_{q}^{\Bar{X}} < 0.5$, which lies below the threshold. Therefore, a repetition code will always reduce the logical bit-flip error rate to some degree. However, as $p_{\Bar{X}} \rightarrow 0.5$, which happens with stronger biasing, more modes are required to gain an appreciable performance improvement. 

While the repetition code has an extremely high bit-flip threshold of $50\%$, it offers zero protection against phase-flips. We have already seen at the GKP level that we cannot completely bias away momentum quadrature displacement errors. There is always a non-zero phase-flip error rate for any finite amount of biasing, potentially resulting in an undetected $\Bar{Z}$ operation. Any even number of phase flips, however, is the product of stabiliser generators and leaves the logical state unchanged, so we only need to be concerned with odd numbers of phase-flips. The number of phase flips~$\abss{\Bar{Z}}$ is also binomially distributed, with $\abss{\Bar{Z}} \sim \mathrm B(n,p_{\Bar{Z}})$. For an $n$-qubit code with an individual phase-flip error rate $p_{\Bar{Z}}$, the probability that an even number of $\Bar{Z}$ errors occurs is given by
\begin{align}\label{eq:phaseflip_rep}
    \Pr(\abss{\Bar{Z}}_{\text{e}})
    \coloneqq
    \Pr\left(\text{$\abss{\Bar{Z}}$ is even}\right)
&=
    \sum_{j=1}^{\floor{n/2}} \binom{n}{2j} p_{\Bar{Z}}^{2j}\left(1-p_{\Bar{Z}}\right)^{n-2j} \nonumber \\
&=    
    \frac{1 + (1 - 2p_{\Bar{Z}})^{n}}{2},
\end{align}
with the phase-flip error rate given by $1 - \Pr(\abss{\Bar{Z}}_{\text{e}})$. As $p_{\Bar{Z}}$ approaches $0.5$ from below (which occurs when $\sigma$ increases in the GDC) or the number of modes, $n$, increases, the success probability $\Pr(\abss{\Bar{Z}}_{\text{e}})$ rapidly approaches $0.5$, which limits the performance of the overall code. Therefore, it is vital to suppress $p_{\Bar{Z}}$ at the GKP level as much as possible. This results in a trade-off in the amount of biasing we choose to apply for our GKP code. We need to bias enough to stop $p_{\Bar{Z}}$ having a significant effect in Eq.~\eqref{eq:phaseflip_rep}, but if we bias too much, we approach the bit-flip threshold, and more modes are required to observe an improvement in Eq.~\eqref{eq:bitflip_rep}. This, in turn, places a stricter requirement on $p_{\Bar{Z}}$. We shall see that optimising the aspect ratio~$r$ is key to observing improved performance with this code. 

Like the GKP layer, the bit-flip [Eq.~\eqref{eq:bitflip_rep}] and phase-flip [Eq.~\eqref{eq:phaseflip_rep}] outcomes are combined to give the overall logical qubit probabilities for the repetition code (hence, `rep', below) with i.i.d. Pauli noise:
\begin{align}
    P_{\text{rep}}[I_L] &= \Pr(\abss{\Bar{X}} \leq k)\Pr(\abss{\Bar{Z}}_{\text{e}}),\label{eq:rep_id}\\
    P_{\text{rep}}[X_L] &= \bigl[1 - \Pr(\abss{\Bar{X}} \leq k)\bigr]\Pr(\abss{\Bar{Z}}_{\text{e}}), \\
    P_{\text{rep}}[Y_L] &= \bigl[1-\Pr(\abss{\Bar{X}} \leq k)\bigr]\bigl[1 - \Pr(\abss{\Bar{Z}}_{\text{e}})\bigr],\\
    P_{\text{rep}}[Z_L] &= \Pr(\abss{\Bar{X}} \leq k)\bigl[1 - \Pr(\abss{\Bar{Z}}_{\text{e}})\bigr].
\end{align}
We now have all the machinery in place to analyse both layers of the biased GKP repetition code. In the next section, we construct the overall code and present key performance results. 

\section{Results \& Discussion}\label{sec:results}
We are now able to assess the overall performance of the biased-GKP-repetition code. The inner rectangular-lattice GKP code takes the continuous GDC noise acting on a single mode and, through the GKP correction process described in Section~\ref{sec:correction_process}, returns the state to the GKP codespace, resulting in either a successful correction or one of $\{\Bar{X},\Bar{Y},\Bar{Z}\}$ being applied to the state. This results in an effective qubit with error rates given by Eq.~\eqref{eq:err} for each of the quadratures. By increasing the lattice aspect ratio~$r$, the $\Bar{Z}$ error rate gets suppressed exponentially at the expense of increasing the $\Bar{X}$ error rate, biasing the qubit. 

The corrected GKP modes are then used as the data qubits in the repetition code. The GKP qubits are assumed to be identical and independent. The error rates calculated from Eq.~\eqref{eq:err} are used as inputs to the repetition code analysis, with $p_{q}^{\Bar{X}}$ mapping to $p_{\Bar{X}}$ in Eq.~\eqref{eq:bitflip_rep} and $p_{p}^{\Bar{Z}}$ mapping to $p_{\Bar{Z}}$ in Eq.~\eqref{eq:phaseflip_rep}. The overall code error rate is then given by $1-P_{\text{rep}}\left[I_{L}\right]$. We denote a concatenated $n$-qubit GKP-repetition code, with GKP biasing~$r$, under the isotropic GDC of variance $\sigma^2$, by $\llbracket n, r, \sigma\rrbracket$.

Using this model, we first discuss the effects of bias~$r$ on the repetition code performance and find that $r$ has to be optimised for each pair $(n,\sigma)$ if one wishes to maximise code performance. We denote a code $\llbracket n, r, \sigma\rrbracket$ with optimised~$r$ by $\llbracket n, \sigma\rrbracket$. We then move to discuss threshold behaviour and find that the biased-GKP-repetition code can suppress errors in the GDC for $\sigma < 0.599$. Finally, we discuss resource scaling for $n$ and~$r$.

\subsection{Bias Optimisation}

\begin{figure}
    \centering
    \includegraphics[width=0.9\linewidth]{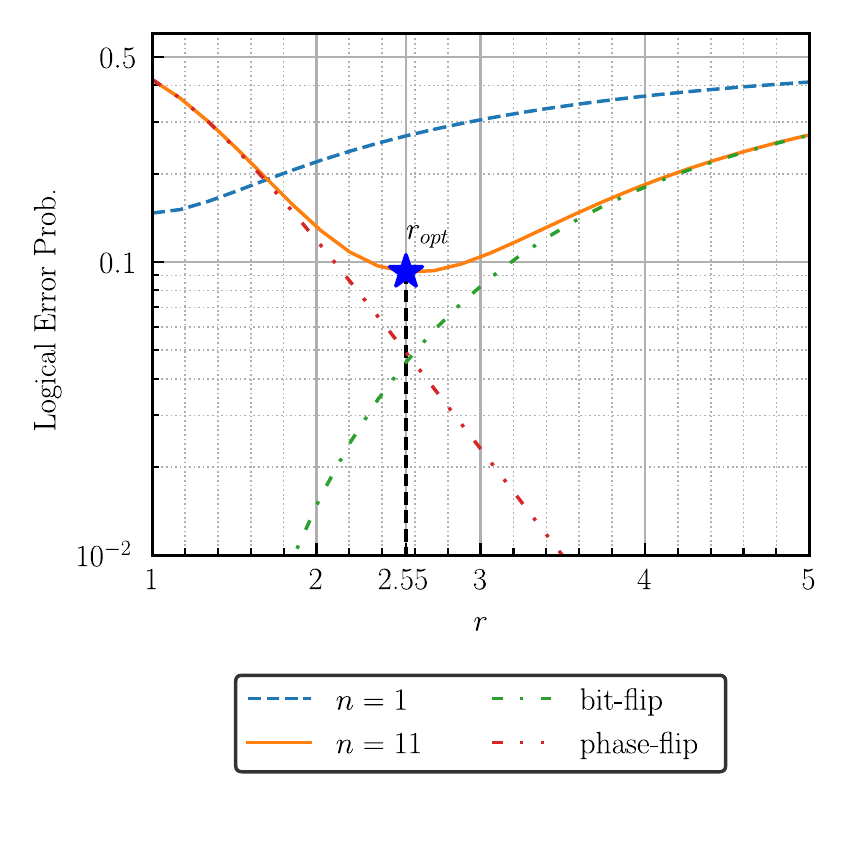}
    \caption{Bias optimisation for the GKP repetition code under a GDC with $\sigma = 0.5$. The solid and dashed lines show overall code error rates given by $1 - P_{\text{rep}}\left[I_L\right]$, using Eq.~\eqref{eq:rep_id}. For a single mode the error rate is minimised at $r = 1$, a square-lattice GKP code. Increasing $n$, we find the performance improves when using rectangular-lattice codes. The bias can be tuned to suit $\sigma$ and number of modes available. If $r < r_{\text{opt}}$, phase-flip errors are present at the GKP level, which impact performance in Eq.~\eqref{eq:phaseflip_rep}, indicated by the double-dot-dashed red line. For $r > r_{\text{opt}}$, the bit-flip errors at the GKP level bring the repetition code closer to the threshold of $0.5$, degrading the performance of Eq.~\eqref{eq:incompletebeta}, which is shown by the increase in the bit-flip error rate for the code in the dot-dashed green line. The performance of the optimised code~$\llbracket {n=11}, {\sigma=0.5}\rrbracket$
    is shown by the blue star and used in the threshold calculations shown in Figure~\ref{fig:threshold_large}. For this example, we find the optimum performance is achieved at $r_{\text{opt}} \approx 2.55$.}
    \label{fig:r_sweep}
\end{figure}

\begin{figure*}[t!]
    \centering
    \includegraphics[width=\linewidth]{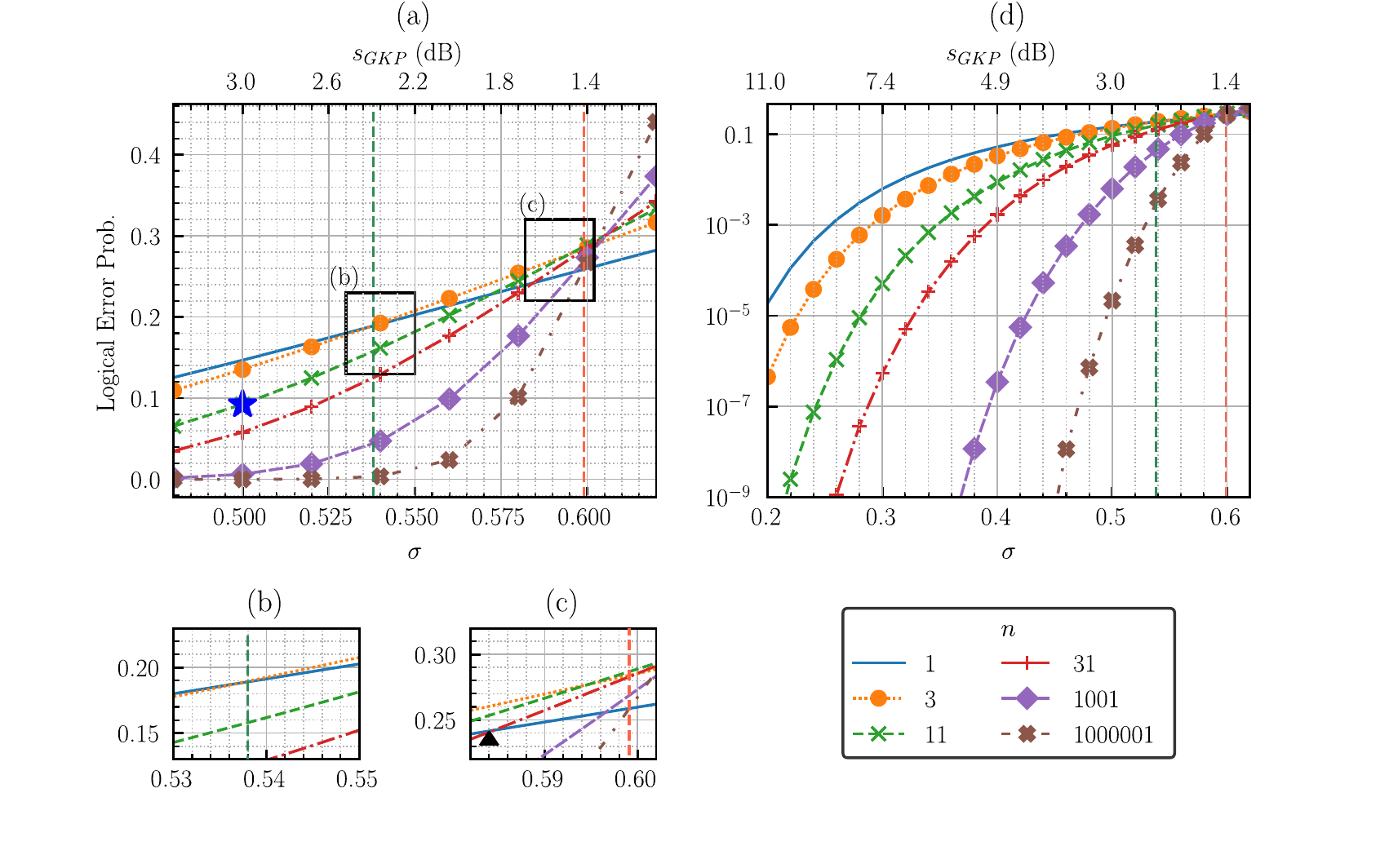}
    \caption{Numerical threshold analysis for the biased GKP repetition code with an estimated threshold of $\sigma \approx 0.599$, highlighted by the vertical dashed red (lighter) line towards the right of the main subfigures. Each point $(\sigma, n)$ selects a code~$\llbracket n, \sigma \rrbracket$ with optimised bias~$r$. The $\llbracket {n=11}, {\sigma=0.5} \rrbracket$ code, used in Figure~\ref{fig:r_sweep}, is indicated by the blue star in~(a). There are three distinct behaviour regimes. For $\sigma < 0.538$, any length repetition code provides a performance increase over a single GKP state, which is shown in~(b) with the crossover point highlighted by the vertical green dashed line. For $0.538 < \sigma < 0.599$, shorter-length codes initially perform worse than a single, square-lattice GKP state, but by increasing $n$, the performance recovers and eventually outperforms the single-state case. This is shown by the crossing point shifting towards higher $\sigma$ for larger $n$ codes. Subfigure~(c) shows an expanded view about the threshold. Beyond this, the code offers no improvement over a single GKP state, and performance degrades with increasing $n$. The black triangle indicates the crossing point of an $n=31$ code, which is located at $\sigma = 0.584$. Subfigure~(d) shows the code performance for a larger range of $\sigma$. Codes of $n \leq 11$ data modes can still provide order-of-magnitude reductions in error rates for modest amounts of noise, such as for $\sigma < 0.3$.
}
    \label{fig:threshold_large}
\end{figure*}

Within the model, we allow~$r$ to vary with the code size and noise. As an example, the performance of an 
$\llbracket {n = 11}, r, {\sigma = 0.5}\rrbracket$
code
under varying bias~$r$ is shown in Figure~\ref{fig:r_sweep}. The performance of a single GKP mode---i.e.,~a $\llbracket {n = 1}, r, {\sigma = 0.5}\rrbracket$ code---is also included for comparison. Changing the bias of the underlying GKP mode has a significant effect on the repetition code performance, with an optimum at $r_{\text{opt}} \approx 2.55$ in this case, and thus $\llbracket {n = 11}, {\sigma = 0.5} \rrbracket$ is shorthand for $\llbracket {n = 11}, {r = r_{\text{opt}}}, {\sigma = 0.5} \rrbracket$ for these values of~$n$ and~$\sigma$. The repetition code's optimum error rate is also below that of the optimum for the single-mode code $\llbracket {n=1}, {\sigma = 0.5}\rrbracket$, which is a square GKP state ($r=1$), indicating that the biased repetition code can effectively suppress noise.

For $r < r_{\text{opt}}$, phase-flip errors have not sufficiently been suppressed at the GKP layer and are still present in the repetition code. Since the code offers no protection against phase-flip errors, we see these dominate the overall error rate, indicated by the red double-dot-dashed line. As the bias is increased, the phase-flip error rate gets exponentially suppressed at the expense of introducing GKP bit-flip errors. Provided the GKP bit-flip rate is below $50\%$, which is always the case for GKP states under the GDC, the repetition code will reduce this rate. However, as $r$ increases, the bit-flip rate approaches the threshold, and the degree of improvement is limited. Bit flips come to dominate the overall performance, as shown by the dot-dashed green line closely matching the overall error rate. For any length code, in the limit of infinite bias, phase-flip errors are completely removed, but $p_{\Bar{X}} \rightarrow 0.5^-$, and the overall error rate tends to $0.5$ from below, shown in Figure~\ref{fig:r_sweep} by the asymptotic behaviour of both the $n=1$ and $n=11$ codes, which approach 0.5 far to the right of the values shown. The value of $r_{\text{opt}}$ varies with both the number of modes and noise. Therefore, for given~$n$ and~$\sigma$, the bias must be re-optimised. The optimum performance, shown in Figure~\ref{fig:r_sweep} by the star, is used for threshold estimation detailed in the next section. 

\subsection{Threshold Behaviour}
A key metric in analysing the performance of a given code family is the threshold, defined as the point beyond which no code in the family can provide a reduction in the error rate compared to an unencoded data qubit. Threshold analysis for the biased repetition code is presented in Figure \ref{fig:threshold_large}, where we estimate a threshold of $\sigma \approx 0.599$. Numerical simulations are performed for $\sigma \in [0.2 ,0.62]$ at $0.02$ intervals, ranging from $n=1$ to $n=10^7$ modes. Around the crossover and threshold points detailed in Figures~\ref{fig:threshold_large}(b) and~(c), simulations are fine-grained at intervals of $0.001$. For each data point, the optimum bias is selected by scanning over the range $r \in [1,15]$, where an example is shown by the star in Figures~\ref{fig:r_sweep} and~\ref{fig:threshold_large}(a) for $\llbracket {n=11}, {\sigma=0.5} \rrbracket$. The ${r=15}$ and ${n=10^7}$ cutoffs are chosen to avoid numerical error.

The code behaviour in Figure~\ref{fig:threshold_large}(a) can be categorised into three distinct regions. For $\sigma < 0.538$, an optimised repetition code of any length performs better than a single GKP mode, with the degree of error suppression increasing monotonically with~$n$. When $0.538 < \sigma < 0.599$, the repetition code initially performs worse for smaller~$n$, shown in~(b), where the ${n=3}$ code now has a higher error rate than the single-mode code. When $n$ is sufficiently large, however, the code will still outperform the single GKP mode. In this region, the code performance still improves with increased~$n$. As the threshold is approached, larger codes are required to get beyond the break-even point and show an improvement, as shown in~(c), where the crossover point moves to the right with increased~$n$. Beyond the estimated threshold of $\sigma \approx 0.599$, shown in~(c), no repetition code does better than the single-mode GKP state. Furthermore, increasing~$n$ now makes the performance worse. While the results indicate that the threshold could be slightly higher than our estimate of $0.599$, we have conservatively taken it as the point in which none of our simulation results outperform a single mode. 

Remarkably, this threshold is higher than the result obtained for the biased-GKP surface code in Ref.~\cite{Hanggli2020}, which quoted a maximum threshold of $\sigma = 0.580$ when decoding the GKP layer without using analogue techniques from Ref.~\cite{Fukui2018b} We cannot directly compare with the biased-GKP XZZX threshold of $\sigma \approx 0.67$ in Ref.~\cite{zhang2023} because the authors only decode using analogue methods. We see no reason why analogue GKP decoding cannot also be employed with the GKP-repetition code to similarly boost the threshold. 

The key difference between our result and that in Ref.~\cite{Hanggli2020} is how bias is employed within the code. For the surface-code result, rectangular-lattice states of fixed~$r$ are employed at the GKP level and then used on a square patch of surface code. The level of biasing remains fixed for all code distances and values of~$\sigma$. In our biased repetition code, we tailor the level of bias for each optimised code~$\llbracket n, \sigma \rrbracket$. We also allow for much higher levels of bias than those used in the surface-code result, where they limited themselves to a maximum of $r=4$. This suggests a trade-off between the two approaches. If one is able to create GKP states with higher levels of bias, then our method can get similar levels of performance to that of a surface code~\cite{Hanggli2020} but, crucially, without the decoder complexity and only needing to employ weight-two stabiliser measurements instead of weight-four.

\subsection{Resource Scaling}

The primary aim of this work is to study the biased-GKP-repetition code without considering resource restrictions to see if the code works in principle. Under these conditions, we have found that our code admits a threshold comparable to the biased-GKP surface code in Ref.~\cite{Hanggli2020}. For the code to be of practical use, however, we need to know how resources scale with the noise level. For the biased repetition code, we have two main resources to consider: the number of modes~$n$ and the amount of biasing~$r$ available per mode. 

Unlike the work completed for the biased-GKP surface code, where $r$ is fixed at regular intervals and limited to a maximum of ${r=4}$, we do not impose any restriction beyond the numerical cutoff at ${r=15}$. The arguments for doing this are twofold.

First, we have found that the optimum value of~$r$ varies with both $\sigma$ and~$n$. Therefore the most advantageous code implementation strategy is to characterise the noise channel, choose the number of data modes available for encoding into the repetition layer, and then generate states at the optimum bias level. 

Second, the ${r=4}$ limitation imposed in Ref.~\cite{Hanggli2020} is based on a maximum GKP squeezing level of $s_{\text{GKP}} = 11~\mathrm{dB}$ which was argued as a realistic target for practical implementations. The $11~\mathrm{dB}$ budget is split into $\approx 4.9~\mathrm{dB}$ for a minimal choice of ${\sigma = 0.4}$ in their simulations and $6~\mathrm{dB}$ corresponding to applying the squeezing operator, Eq.~\eqref{eq:squeeze_operator}, with $r=4$ to a square-lattice. This interpretation corresponds, however, with the biasing being implemented at the channel level and not through the lattice asymmetry discussed in Section~\ref{sec:noise_model}. That is, the noise spikes of the Wigner function in Ref.~\cite{Hanggli2020} are distributed on a square lattice and have asymmetric variance.

In contrast, we are subjecting an ideal rectangular-lattice GKP state to an isotropic GDC, where the noise spikes in both quadratures have an equal variance of $\sigma^2$ that is independent of the biasing $r$. The biasing of the logical GKP qubit comes solely from the spacing of the lattice and not the shape of the noise distribution. Restricting ourselves to the same $11~\mathrm{dB}$ limit corresponds to the minimum noise of ${\sigma = 0.2}$ employed in our analysis.

\begin{figure}[t!]
    \centering
    \includegraphics[width=0.95\linewidth]{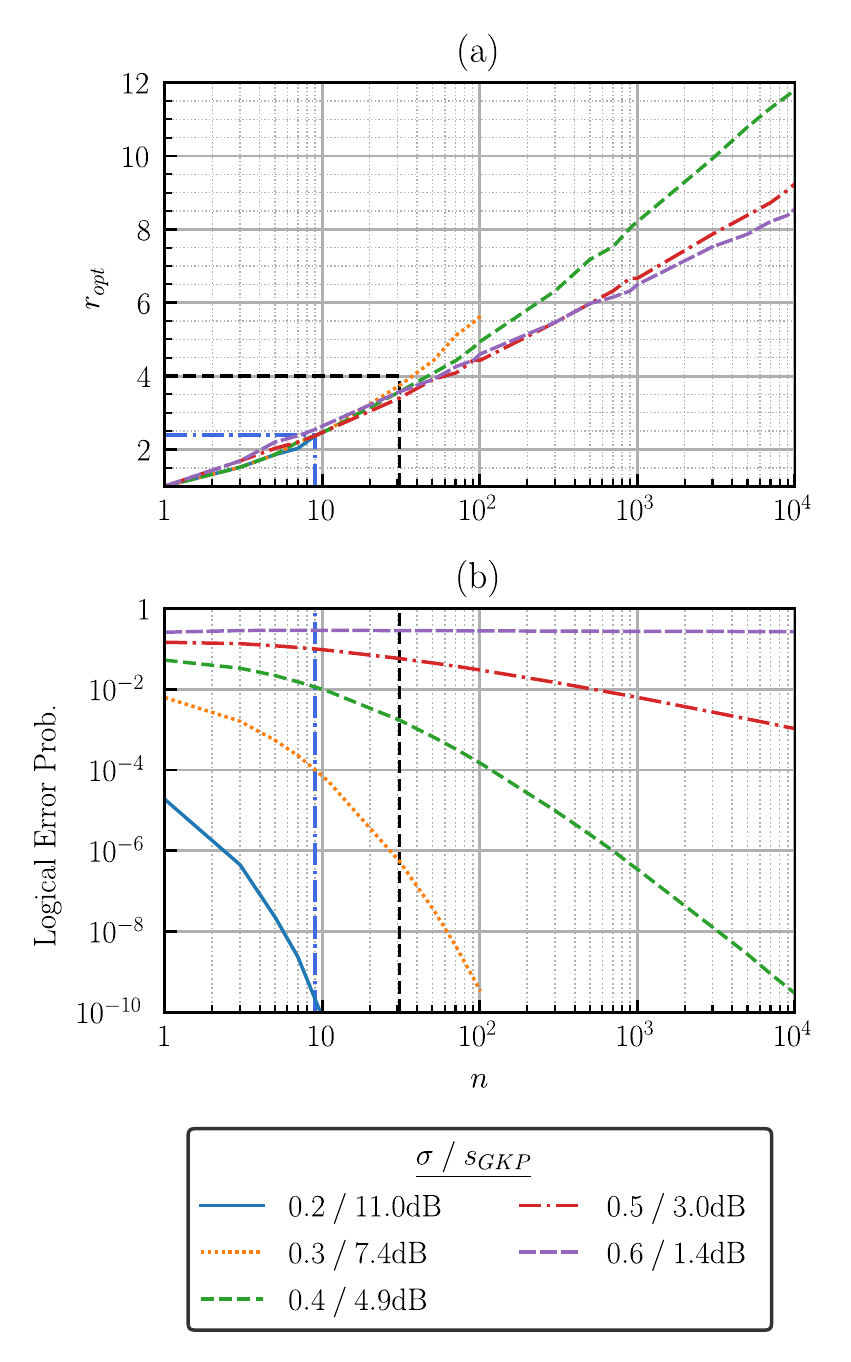}
    \caption{Resource Scaling for the biased GKP repetition code. (a) shows how the optimum bias $r_{\text{opt}}$ varies with $n$ and $\sigma$. The required bias increases approximately linearly with an exponential increase in the code length. For smaller codes the noise severity has minimal effect on the optimum bias level, however as the code scales this becomes more significant as seen by the divergence of the plots. For fixed $n$, the optimum bias is actually reduced for increased $\sigma$. Enforcing an upper limit on the biasing $r$ restricts the maximum $n$ code that can be utilised. For example, limiting to $r \leq 4$---to compare with \cite{Hanggli2020}---only allows for codes of length up to $n \approx 31$ to be used at their optimum bias, indicated here by the black dashed line. (b) shows how logical error rates scale when increasing the code length $n$. The scaling becomes more demanding with increased $\sigma$ until the threshold, beyond which the code offers no reduction in the error rate. Limiting $n$, whether due to finite biasing or size of system places a ceiling on the degree of error suppression the code can offer. However, for a handful of modes ($n\leq 9$, indicated by the blue dot-dashed line) error rates can still be significantly suppressed providing initial noise levels are low enough---$\sigma \leq 0.3$ in this case.}
    \label{fig:scaling}
\end{figure}

The question remains open whether physical rectangular GKP states are intrinsically more difficult to generate than their square-lattice counterparts. Modifying existing generation protocols to include rectangular states and assessing their performance is suggested for future work. Assuming biased states are harder to produce, placing an upper limit on the level of bias available restricts the number of modes that can be effectively used in the repetition code, as shown in Figure~\ref{fig:scaling}(a). We see that as the code length~$n$ increases, a higher optimum bias~$r_{\text{opt}}$ is required to effectively suppress error rates. If~$r$ is restricted and we continue to increase~$n$, we will sit below the optimum shown in Figure~\ref{fig:r_sweep}, where phase-flip errors dominate the overall performance. Counterintuitively, increasing~$\sigma$ slightly reduces the bias requirements, but this doesn't become significant until large numbers of modes are used. Restricting to a maximum of $r=4$ for comparison with the surface-code result~\cite{Hanggli2020}, we see that only codes up to ${n \approx 31}$ can be used at the optimum bias level, shown by the black dashed cutoff. This limit on the code length in turn caps the performance. Figure~\ref{fig:scaling}(b) shows logical errors rates with increasing~$n$ for a selection of noise levels, corresponding to vertical slices of Figure~\ref{fig:threshold_large}(d). As~$\sigma$ increases, more modes are required to effectively suppress the errors shown by the reduced gradient until the threshold is passed where no value of~$n$ yields a performance improvement. The cutoff limits us to the region left of the dashed line, meaning arbitrarily small error rates can no longer be achieved with the repetition code. However, significant reductions in error rates can still be achieved with these codes, as shown by the multiple orders of magnitude of improvements shown for $\sigma \leq 0.4$.

By restricting the maximum level of biasing available, we are effectively capping the code length $n$ available to us. This has two main effects on the overall code performance. First, the break-even point, where the repetition code crosses from performing better to worse compared to a single GKP mode, is reduced. For the $n=31$ code the crossing point is located at $\sigma = 0.584$, shown by the black triangle in Figure~\ref{fig:threshold_large}(c). This is still higher than the surface-code result~\cite{Hanggli2020} of $\sigma = 0.580$, but the code family now no longer exhibits threshold behaviour, as errors can no longer be arbitrarily suppressed below the break-even point. Lack of a threshold is the second result of limiting the bias. Therefore, we conclude that the effectiveness of the biased repetition code will ultimately depend on the levels of biasing that can be generated in physical GKP states. 

The upper limit on~$r$ significantly impacts the power of the biased repetition code around the threshold. Significant improvements can still be achieved, however, for moderate levels of biasing and noise. For example, consider noise level of ${\sigma = 0.3}$~/~$ 7.4~\mathrm{dB}$. With only 9 data modes and a maximum bias of~${r=2.4}$, shown by the blue dashed line in Figure~\ref{fig:scaling}, the logical error rate gets suppressed to approximately~$10^{-4}$, a 60-fold improvement over the single GKP mode. The logical mode has an error rate comparable to that of an 8.7~dB square GKP mode. Furthermore, if we allow ourselves ${r=4}$ states then choosing ${n=31}$ improves the noise levels of a 7.4~dB channel to that of a 12.3~dB square-lattice state, which meets fault-tolerance thresholds for existing GKP qubit concatenation schemes~\cite{Noh_2022}, opening up the potential for using the biased-GKP repetition code as a low-level encoding scheme. This is all achieved with a fixed linear overhead, weight-two stabiliser operators, and parity checks at the repetition level.

\subsection{Open Challenges}
\label{subsec:challenges}

While promising, we reiterate that all results have been obtained studying code-capacity noise where all encoded states, operations, and measurements have been assumed to be ideal. To ensure the code can provide a practical benefit, analysis needs to be extended to account for imperfections throughout the correction process  and be extended to more realistic noise models that more closely resemble physical implementations.  Furthermore, this work has only studied the performance of the code as a quantum memory. This needs to be extended to performing universal logic on the encoded states. In such a context, bias preservation under gate operations will be a challenge. The below challenges we identify hold for any code that hopes to use biased-GKP states as a bosonic encoding.

The immediate challenge to address is the generation of physical biased GKP states. For ideal states, rectangular GKP $\ket{\Bar{0}}$ and $\ket{\Bar{+}}$ states can be obtained by acting with the squeezing operator, Eq.~\eqref{eq:squeeze_operator}, on a square-lattice state. However, this does not hold in the cases where the state is physical or suffering from GDC noise. In both of these cases, the Wigner peaks have been blurred to Gaussian distributions, so squeezing will both change the lattice spacing---which is desired---and squeeze the individual peaks as an unwanted side effect. Any GKP generation scheme will need to directly produce approximate rectangular states to see any advantage. It is assumed that directly generating rectangular states will be more challenging than generating those with a square lattice. Modifying existing GKP state-generation proposals and assessing their practicality for rectangular states is highlighted for potential future work.

 Accounting for physical rectangular GKP states will naturally reduce the code performance under a Gaussian displacement channel, as the Gaussian noise from the physical state and channel are additive. However, if the total error variance of the channel and state is still below threshold, we anticipate that the biased GKP-repetition code will still be able to suppress logical errors.

Here, we provide some intuition as to what we expect under the more physically relevant case of photon loss. The effects of the photon-loss channel are twofold. First, each peak in the GKP state gets blurred just as a displacement channel. Second, the peaks are all pulled in towards the origin, with peaks further from the origin suffering larger shifts~\cite{Terhal2016}. Amplifying the state either before or after the loss results effectively in a Gaussian displacement channel~\cite{Albert2018, Noh2019}, and more clever ways exist to deal with this noise in certain cases, including simply rescaling measurement outcomes under certain conditions~\cite{Fukui2020}. Since loss can be reduced to the GDC (at the expense of extra noise), this further justifies our focus on the latter noise model. We leave a detailed analysis of loss and other physically relevant channels to future work.

Once rectangular-lattice states have been generated, the next challenge to address is implementing logical operations at the GKP level that preserve the noise bias. This will be needed for any physical implementation of the GKP-repetition, biased GKP-Surface \cite{Hanggli2020}, and biased GKP-XZZX \cite{zhang2023} codes, including quantum memory experiments. Furthermore, to provide a fair comparison to other schemes, such as the cat-repetition code \cite{Guillaud2019,Guillaud2021} which is analysed in the finite energy regime, a full `circuit-level' analysis of the code, where the whole correction circuit is simulated, is required. This in turn requires a set of bias preserving gates. Here, we give some thoughts about these considerations, while leaving a detailed analysis to future work.

One advantage of the GKP code is that all encoded logical Clifford gates can be realised with Gaussian operations. This remains true for any lattice choice. For a square-lattice GKP code, we can implement all single-mode Clifford gates using $\{{\Bar{Z} = \op{Z}(\sqrt{\pi})},{\Bar{H} = \hat{R}(\pi/2)}, {\Bar{S} = \op{P}(1)}\}$. Only the encoded~$\Bar{S}$ gate requires squeezing through the unit shear operator. With a rectangular code, the logical Hadamard becomes
\begin{align}
    \Bar{H} &= \hat{S}(\sqrt{r})\hat{R}(\pi/2)\hat{S}(\sqrt{r^{-1}})\\
    &= \hat{S}(\sqrt{r})^{2}\hat{R}(\pi/2)\\
    &= \hat{S}(r)\hat{R}(\pi/2),\label{eq:logichad}
\end{align}
where $\hat{S}(\sqrt{r^{-1}})$ first converts the rectangular state to the square-lattice encoding, $\hat{R}(\pi/2)$ applies the square-lattice Hadamard, and $\hat{S}(\sqrt{r})$ re-encodes the state back into the rectangular-lattice code. The standard Fourier rotation is recovered for ${r=1}$. On the second line we pulled the squeezing operation through the $\pi / 2$ rotation, which has the effect of rotating the squeezing axis by $\pi / 2$ also. This combines with the original squeezer after the rotation, which results in a total squeezer that converts from the rotated rectangular-lattice code (with bias~$r^{-1}$) back to the original rectangular-lattice code (with bias~$r$). This is another way to view the action of this gate. 

Therefore, to implement a logical Hadamard for a rectangular state with bias~$r$, total squeezing by a factor of~$r$ is required. Inline squeezing can be avoided by using a measurement-based model, in which any Gaussian operation can be realised by choosing appropriate homodyne measurements, provided the squeezing of the cluster-state nodes is high enough~\cite{Menicucci2006,Gu2009,Walshe2020}. Furthermore, to extend to universality at the GKP level, an encoded~$\Bar{T}$ gate is required. For the square- and hexagonal-lattice codes, this can be achieved by distilling magic states through error correcting the vacuum~\cite{Baragiola2019}. These results are straightforwardly extendable to the rectangular-lattice case using the methods described in that reference.

To take advantage of the biased nature of the code one needs to ensure that the GKP logical gate set is itself bias preserving. Without this, noise can be transferred from the position quadrature, which is protected by the repetition code, to the momentum quadrature which is completely unprotected. When the rectangular-lattice logical Hadamard, Eq.~\eqref{eq:logichad}, acts on a noisy GKP state, it does exactly that through its combination of rotation and squeezing. Therefore, the standard Clifford plus T gate-set - generated by $\{\Bar{H},\Bar{S},\Bar{T}\}$ - can no longer be used in the presence of biased noise and an alternative must be found. This has been achieved for physical \cite{Aliferis2008,Webster2015} and cat qubits \cite{Guillaud2019,Guillaud2021}. Finding a universal bias-preserving gate set for GKP qubits remains an open question.

Finally, once we have rectangular-GKP states and bias-preserving operations, we can consider implementing logic at the repetition-code level. A consequence of the Eastin-Knill theorem~\cite{Eastin2009} is that for a qubit error-correcting code, one cannot realise a universal encoded gateset transversally. For an $n$-qubit bit-flip repetition code, the encoded Pauli operators are given by
\begin{align}
    X_L &= \Bar{X}^{\otimes n},\\
    Z_L &= \Bar{Z}_{i} \quad \text{for any $i \in \{1,\dotsc,n\}$}.
\end{align}
Keep in mind that these can also be multiplied by any stabiliser to give another logical operator that does the same thing---e.g.,~the $Z_L$ operation can also be implemented by applying $\Bar{Z}$ on any odd number of qubits.

These logical gates are transversal. Furthermore, any phase rotation can be implemented by just acting on a single data qubit, so the $S_L$ and $T_L$ gates can also be implemented transversally. Therefore, by Eastin-Knill the $H_L$ gate will not be transversal. Finding a full, bias-preserving gate set to upgrade the code from a quantum memory is left for future work.  

\section{Conclusions \& Future Outlook}\label{sec:conclusions}
In this work, we have introduced the biased-GKP-repetition code and studied its performance under the isotropic Gaussian displacement channel~(GDC) in the code-capacity model. The inner layer of the code consists of a rectangular-lattice GKP code that both discritises and biases the continuous noise from the GDC to suppress GKP~$\Bar{Z}$ phase-flip errors. Concatenating with a bit-flip repetition code then allows for an overall reduction in the logical error rate. Under this model, we numerically estimate a threshold of ${\sigma \approx 0.599}$ which outperforms the biased-GKP surface code in Ref.~\cite{Hanggli2020} at the expense of increased biasing. A key advantage of this code is that all stabiliser measurements at the qubit level are only weight two, as opposed to weight 4 for any surface code variety. Furthermore the decoding complexity is drastically reduced from that of the surface code by employing a simple repetition code. 

We find that the main obstacle to effectively implementing this code as a quantum memory will be the degree of biasing available at the GKP level. The biasing level available limits the largest-length repetition code that can be used effectively. This, in turn, constrains the levels of error suppression available. Even with a moderate amount of biasing and just a handful of modes, however, significant reductions in error rates can still be achieved for channels of ${\sigma < 0.3}$ which corresponds to $>7.4~\mathrm{dB}$ of GKP squeezing, opening up the possibility for using this code as a low-level encoding for future fault-tolerance schemes. 

We also identify and outline a number of challenges to implementing the biased-GKP-repetition code that also hold for other proposals of concatenated rectangular-lattice GKP codes, such as those in Refs.~\cite{Hanggli2020,zhang2023}. These are mainly the generation of rectangular-lattice states, assessing the impact of using physical approximate GKP states, and implementing bias-preserving logical gates at the GKP level, all of which are suggested for future work.

\acknowledgments
We thank Rafael Alexander, Ben Baragiola, and Lucky Antonopoulos for useful discussions. This work is supported by the Australian Research Council via the Centre of Excellence for Quantum Computation and Communication Technology (CQC$^2$T) (Project No.\ CE170100012). MS acknowledges support from EPSRC Quantum Engineering Centre for Doctoral Training EP/S023607/1 and studentship support from the European Research Council starting grant PEQEM ERC-2018-STG 803665.

\bibliography{biasedGKPrefs.bib}

\end{document}